# Comparison of Multiplexing Policies for FPS Games in terms of Subjective Quality

Jose Saldana, Julián Fernández-Navajas, José Ruiz-Mas, Luis Sequeira, Luis Casadesus [1]

*Abstract*—This paper compares two policies which can be used for multiplexing the traffic of a number of players of a First Person Shooter game. A network scenario in which a number of players share an access network has been simulated, in order to compare the policies in terms of a subjective quality estimator. The first policy, namely *timeout*, achieves higher bandwidth savings, while the second one, *period*, introduces less delay and jitter. The results show that the difference in terms of QoE is only significant when the number of players is small. Thus, in order to make the correct decision, the concrete network scenario and the characteristics of the router would have to be considered in each case, taking into account the estimation of the subjective quality that can be expected.

*Index Terms*— First Person Shooter, multiplexing, subjective quality, online gaming, QoE

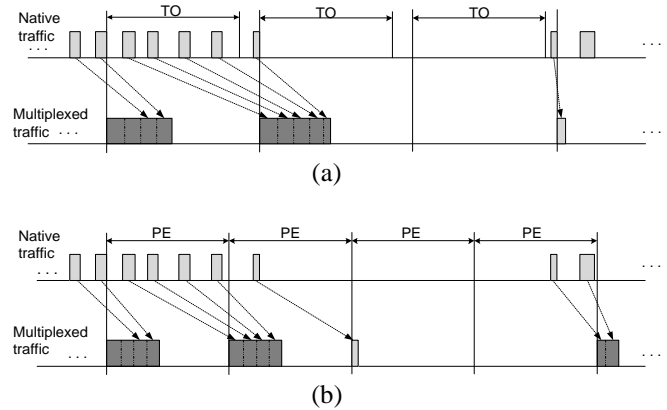

Figure 1. Behaviour of *timeout* (a) and *period* (b) policies.

## I. INTRODUCTION AND RELATED WORKS

FIRST Person Shooter (FPS) games have been reported as the ones with the tightest real-time requirements. The companies that provide them have to deploy network infrastructures capable of supporting the service with very low latencies. The network objective impairments that affect the QoS are mainly delay, jitter and packet loss.

Multiplexing is a well-known concept, which can be used so as to increase the bandwidth efficiency of real-time services, thus reducing network impairments that affect QoS. It can be applied when many flows share the same path, by merging a number of packets from different flows into a bigger multiplexed packet, as it happens with VoIP [1].

Multiplexing has been adapted to the traffic of FPS games, using different policies to select the native packets to be included into a multiplexed one [2]. As a summary of the multiplexing method, it can be said that it first defines an interval, and when it expires all the received packets, with their headers compressed, are multiplexed with PPPMux, and sent using an L2TP tunnel. The method achieves significant bandwidth savings when applied to client-to-server traffic, so we will use it in our tests.

The difference between the two policies (Fig. 1), which are named *timeout* (a) and *period* (b), is that the former checks if the interval has finished each time a packet is received, while the latter sends a packet exactly when the interval ends. There are some exceptions: if a single packet has arrived, it is sent in its native form, in order to avoid the increase of the overhead; if a size next to MTU (Maximum Transmission Unit, 1,500 bytes) is achieved, then the multiplexed packet is sent immediately. In our tests, we have used the threshold value of 1,350 bytes: if a packet arrives and this size is reached, then all the retained packets are immediately sent.

In [2] a comparison between the policies was performed using the traffic of the FPS game *Half Life Counter Strike 1*. The results showed that, although *timeout* policy is able to achieve higher bandwidth savings, it adds more jitter, since in this case multiplexing delay does not have an upper bound. On the other hand, *period* policy introduces a uniformly distributed delay, but its bandwidth saving is smaller.

But considering QoS parameters does not suffice for service providers, which want to know the user's perception of the offered service. So they may prefer tools that, with a reasonable accuracy, estimate the perceived quality, normally measured as a Mean Opinion Score (MOS). QoE estimators have been developed for many services [3], and also for some FPS games [4], [5].

In this paper we will compare these two multiplexing policies in terms of QoE, by means of a subjective quality estimator. The effect of the access and transport networks on the delay will also be considered, because the traffic is modified when multiplexing, i.e. a new delay and more jitter are added, and packet size is increased as well.

The rest of the paper is organized as follows: the next section details the test methodology; section III shows the results, and the paper ends with the conclusions.

[1] Communication Technologies Group (GTC) – Aragon Inst. of Engineering Research (I3A) Dpt. IEC. Ada Byron Building. CPS Univ. Zaragoza, 50018 Zaragoza, Spain, e-mail: {jsaldana, navajas, jruiz, lsequeirav, luis.casadesus} @unizar.es

## II. TEST METHODOLOGY

We have selected G-Model [4], which was the first subjective quality estimator developed for a FPS, namely *Quake IV*. It is able to calculate a MOS (between 1 and 5) based on delay and jitter. Delay is measured as Round Trip Time (RTT), and jitter is measured as the standard deviation of the delay. Packet loss is not considered (unless it exceeds 35%), as the game implements a very efficient packet loss concealment algorithm, as reported in [6]. Although MOS is normally considered acceptable for values above 3.5, some studies consider that a value of 3 can also be acceptable for many users [7].

The conducted tests follow the scheme of Fig. 2. We can see that the total delay for client-to-server packets can be calculated as the sum of three components: first, multiplexing delay, which is the time the packets are retained in the multiplexer, waiting for the interval to finish. Second, there is a delay caused by the buffer of the router. And third, the network delay. As found in [8], the covariance between multiplexing and router delays is negative, so they cannot be considered as independent. In fact, there is a mutual relationship: if the multiplexing interval changes, then the total traffic offered to the router will be modified, and so will be the queuing delay. On the other hand, we will consider network delay as independent from other delays.

The traffic generated by this game is a typical example of FPS traffic: a big amount of packets per second (64 pps average), using small sizes (79.5 bytes on average). *Quake IV* traffic traces have been obtained from the CAIA project [9], and they have been properly merged, as done in [2], in order to obtain the traffic of different numbers of players.

The size of the router buffer has been set to 10 kB: as seen in [8], the so-called *tiny* buffers, proposed in [10], present a better behavior in this scenario, although they may increase packet loss. Game packets share the buffer with a variable amount of background traffic, which has been modeled using the following size distribution: 50% of the packets are of 40 bytes, 10% are of 576 bytes, and 40% are of 1,500 bytes [11].

The uplink bandwidth of the access network is set to 2 Mbps. In order to calculate each point of the graphs, 400 sec. of traffic have been simulated in Matlab. Network delays are added offline following the distribution proposed in [12], which divides the delay into a fixed one caused by geographical distance, and a lognormal one, produced by the variable nature of network traffic. The used values are 15 ms for the fixed delay, considering an intra-region scenario, and a lognormal one of 15 ms with a variance of 5.

## III. TESTS AND RESULTS

As we have said, G-Model is based on delay and jitter. Fig. 3 shows the average multiplexing delay, which is roughly half the *period* or the *timeout*. It can be seen that the delay introduced by *timeout* policy is slightly higher. We can also obtain a first conclusion: since this game generates a big amount of packets per second, interval values above 25 ms are not useful, because the threshold size is achieved, triggering the sending of a multiplexed packet before the end of the interval.

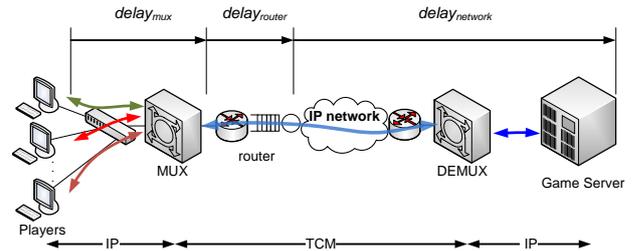

Figure 2. The different delays that affect the traffic

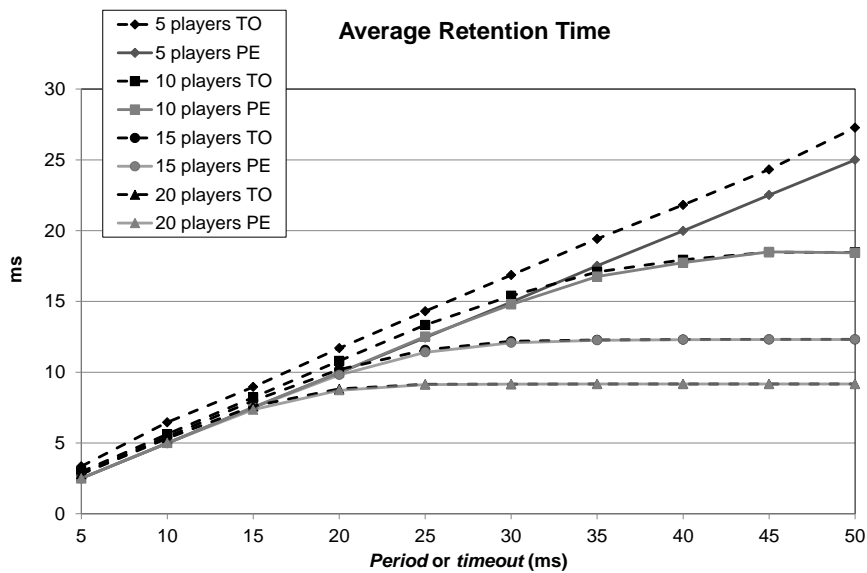

Figure 3. Average multiplexing delay for *period* and *timeout* policies

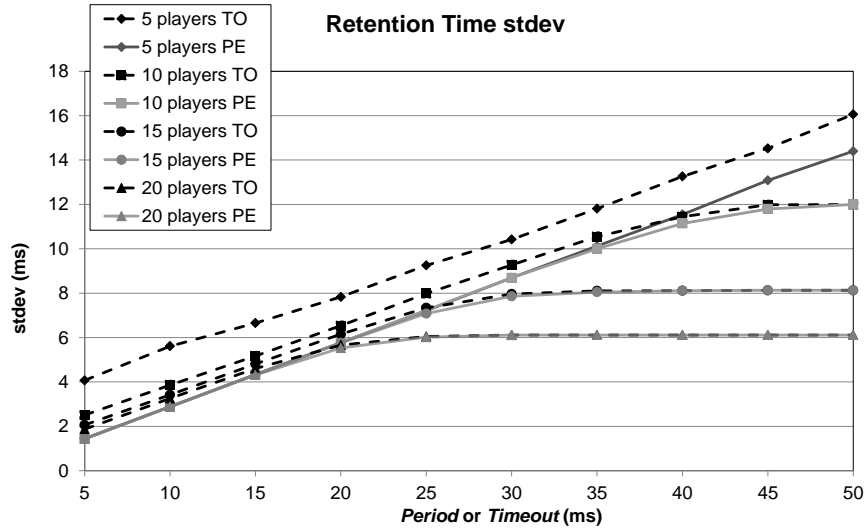

Figure 4. Standard deviation of retention time using *period* and *timeout* policies

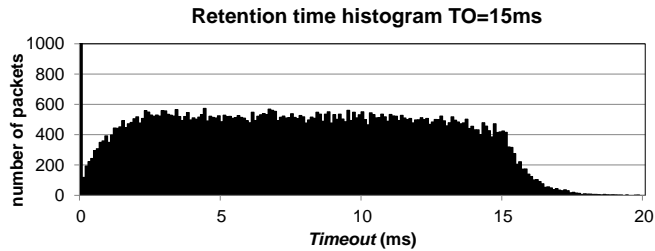

Figure 5. Retention time histogram for *timeout* policy and 20 players. *TO*=15ms. A peak of 4,119 packets with delay=0 has been cut out for clarity

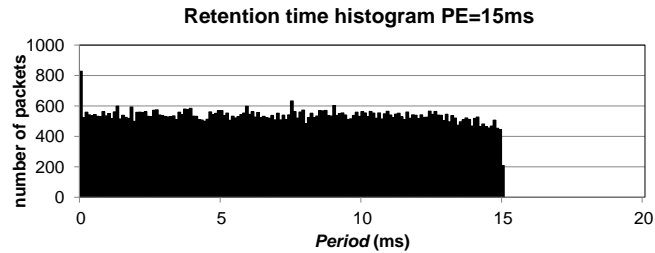

Figure 6. Retention time histogram for *period* policy and 20 players. *PE*=15ms.

Fig. 4 shows the standard deviation of the retention delay. In order to better understand this figure, the retention time histograms for both policies are also shown: the *timeout* histogram (Fig. 5) presents a tail above 15 ms, as the added delay lacks an upper bound. This implies a higher standard deviation than the one for *period* policy (Fig. 6). The peak for delay = 0 in Fig. 5 corresponds to the packets that arrive after the *timeout* and trigger the sending of a multiplexed packet. The peak of roughly 800 packets in Fig. 6 corresponds to the packets that fill the size of 1,350 bytes.

The next graphs show the results for the whole scenario, including access and network delays: Fig. 7 shows the delay, jitter and MOS when the traffic of 5 players is multiplexed. Three different multiplexing intervals have been used: 5, 15 and 25 ms. We can see that, as expected, *timeout* policy adds a higher delay (Fig. 7a). Although this policy achieves a slightly higher bandwidth saving, consequently reducing queuing time, its global delay is higher, because of the higher retention time. The behavior of the jitter is similar (Fig. 7b): *period* policy presents a smaller delay standard deviation. As a result, when MOS is calculated (Fig. 7c) from delay and jitter, *period* policy achieves better results.

The MOS differences between *period* and *timeout* policies, with different numbers of players using an interval of 5 ms, and 1,000 kbps of background traffic, have been included in Table I.

On the one hand, if we only considered bandwidth saving, then *timeout* policy would have been selected. On the other hand, if delay and jitter had been the most important parameters, then *period* policy would have been considered the best one. So we have to merge all the parameters of the whole scenario, using a subjective quality MOS, in order to make a good decision depending on the concrete situation.

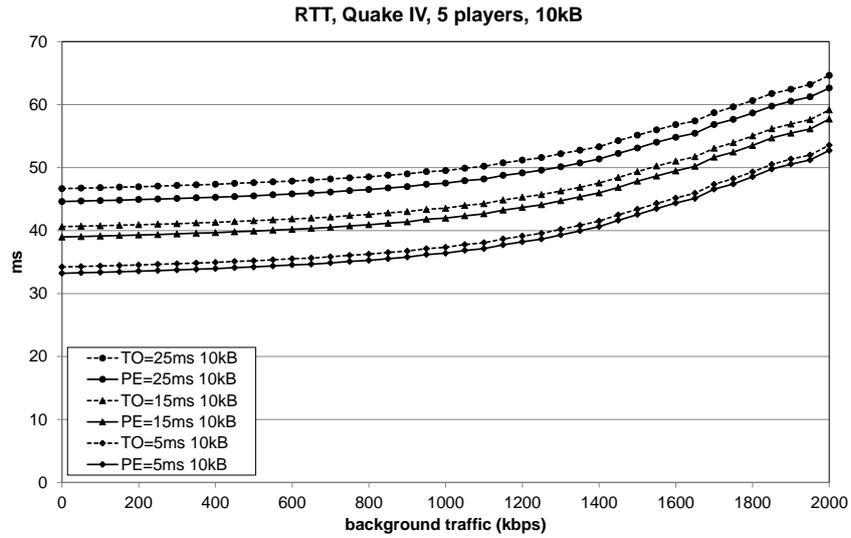

(a)

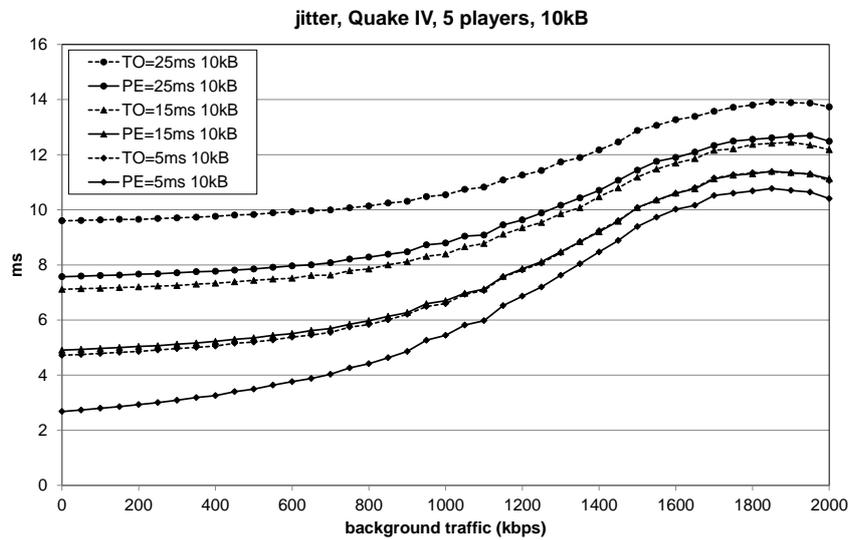

(b)

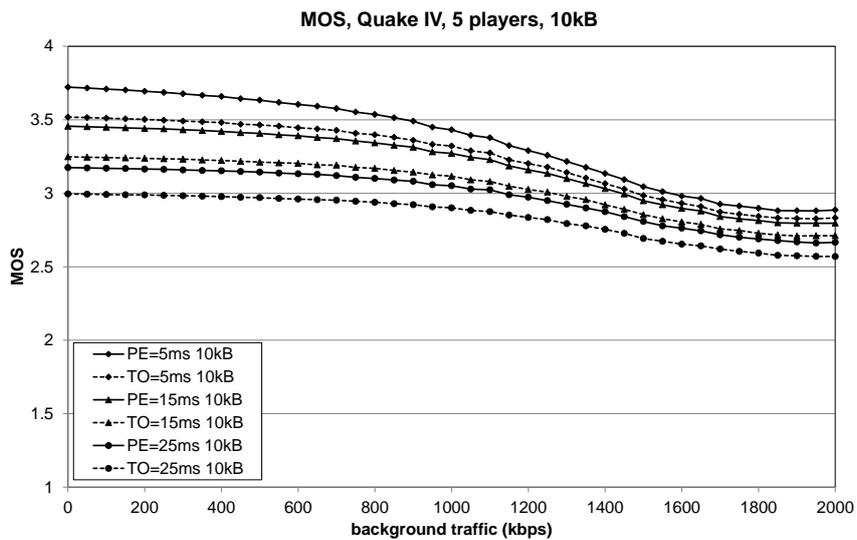

(c)

Figure 7. a) RTT, b) jitter (delay stdev) and c) G-Model MOS for 5 players, a router buffer of 10 kB and different multiplexing intervals

TABLE I. MOS DIFFERENCES USING *PERIOD* AND *TIMEOUT* POLICIES

| Number of players | $MOS_{period}$ | $MOS_{timeout}$ | Difference (%) |
|---|---|---|---|
| 5 | 3.43 | 3.32 | 3.31 % |
| 10 | 3.37 | 3.34 | 0.98 % |
| 15 | 3.30 | 3.28 | 0.42 % |
| 20 | 3.19 | 3.19 | 0.10 % |

Finally, it can be observed that the difference between the two policies can only be appreciated when the number of players is small. As the number of players grows, the advantage of *period* policy is reduced, so the behavior of the two policies becomes very similar.

## IV. CONCLUSION

In this paper, two different policies for online games traffic multiplexing have been compared in terms of a subjective quality estimator. The first one, namely *timeout* policy, achieves better bandwidth savings. The second one adds less delay and jitter. So all the parameters of the scenario have been integrated in a subjective quality MOS, in order to make a good decision. The MOS results show a slight advantage when *period* policy is used. But this advantage gets reduced as the number of players grows. The higher bandwidth saving of *timeout* policy is not translated into a smaller delay, because the time spent at the buffer router is not significantly reduced.

The aim of this paper was to help us to make the correct decision of which is the best multiplexing policy to use. The similarity of the results makes us conclude that the concrete problems of each network will determine the decision, and so will the implementation of the multiplexer, i.e. *timeout* policy may be easier to implement in a low-end machine, as it does not have to perform an active waiting, since the multiplexing mechanism is only activated when a packet arrives. On the other hand, if the multiplexer is included as a process into a computer with enough processing capacity, *period* policy will be more convenient.

ACKNOWLEDGMENT

This work has been partially financed by CPUFLIPI Project (MICINN TIN2010-17298), MBACToIP Project, of Aragon I+D Agency and Ibercaja Obra Social, Red Temática en Codificación y Transmisión de Contenidos Multimedia (TEC2010-11776-E), and NDCIPI-QQoE Project of Catedra Telefonica, Univ. of Zaragoza.